\documentclass[twocolumn,showpacs,preprintnumbers,amsmath,amssymb,
superscriptaddress,groupedaddress]{revtex4}

\usepackage{graphicx}
\usepackage{dcolumn}
\usepackage{bm}

\begin{document}

\title{Statistics of wave interactions in nonlinear disordered systems}%
\author{D. O. Krimer}
\affiliation{Max Planck Institute for the Physics of Complex Systems, 
N\"othnitzer Str. 38, D-01187 Dresden, Germany}
\author{S. Flach}
\affiliation{Max Planck Institute for the Physics of Complex Systems, 
N\"othnitzer Str. 38, D-01187 Dresden, Germany}

\date{\today}

\begin{abstract}
We study the properties of mode-mode interactions for waves propagating in nonlinear disordered one-dimensional systems. We focus on i) the localization volume of a mode which defines the number of interacting partner modes, ii) the overlap integrals which determine the interaction strength, iii) the average spacing between eigenvalues of interacting modes, which sets a scale for the nonlinearity strength, and iv) resonance probabilities of interacting modes. Our results are discussed in the light of recent studies on spreading of wave packets in disordered nonlinear systems, and are related to the quantum many body problem in a random chain.
\end{abstract}

\pacs{05.45-a, 05.60Cd, 63.20Pw}
\maketitle

\section{Introduction}

In the absence of nonlinearity (or many-body interactions in quantum systems) all eigenstates in one-dimensional random lattices with disorder are spatially localized. This is Anderson localization \cite{PWA58}, which has been discovered fifty years ago in disordered crystals as a localization of electronic wavefunction. It can be interpreted as an interference effect between multiple scatterings of the electron on random defects of the potential. Recent experiments on the observation of Anderson localization were performed with light propagation in spatially random optical media \cite{Exp, Exp2}, with noninteracting Bose-Einstein condensates expanding in random optical potentials \cite{Billy,Roati},   and with wave localization in a microwave cavity filled with randomly distributed scatterers \cite{microwave}.

In many situations nonlinear terms in the wave equations (respectively, many body interaction terms in quantum systems) have to be included. Thus, a fundamental question which has attracted the attention of many researchers is what happens to an initial excitation of arbitrary shape in a nonlinear disordered lattice. 
Nonlinearity renormalizes excitation frequencies, thereby inducing interaction between NMs. Numerical studies show that wave packets spread subdiffusively and Anderson localization is destroyed \cite{PS08,fks09,skfk09,lbksf10}. In the regime of strong nonlinearity, far from where it can be treated perturbatively, new localization effects of selftrapping occur \cite{KKFA08}. A theoretical explanation of the subdiffusive spreading was offered in Refs. \cite{fks09,skfk09,F10}.  It is based on the fact that the considered models are in general nonintegrable. Therefore deterministic chaos will lead to an incoherent spreading. Estimates of the excitation transfer rate across the packet tail are obtained by calculating  probabilities of mode-mode resonances inside the packet. Some predictions of this approach include the effect of different degrees of nonlinearity and were successfully tested in \cite{SF10}. 

In this work we study the statistical properties of mode-mode interactions. We focus on i) the localization volume of a mode which defines the number of interacting partner modes, ii) the overlap integrals
which determine the interaction strength, iii) the average spacing between eigenvalues of interacting modes which sets a scale for the nonlinearity strength, and iv) resonance probabilities of interacting modes.
We discuss the results in the light of recent studies \cite{fks09,PS08,skfk09,KKFA08,F10,SF10} on spreading of wave packets in disordered nonlinear systems, and relate our findings to the quantum two interacting particle problem in a random chain.

\vspace{-0.5cm}

\begin{figure}
\includegraphics[angle=0,width=0.8\columnwidth]{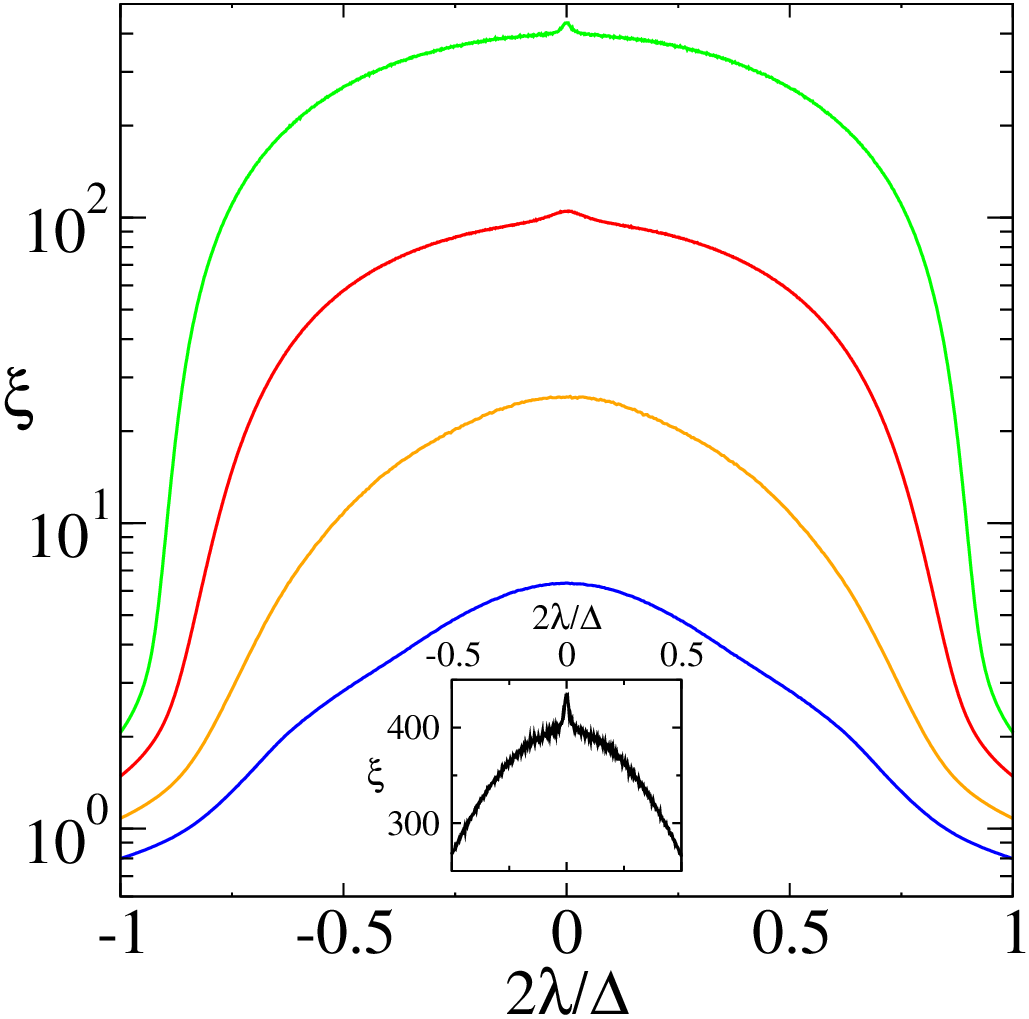}
\caption{(Color online) Localization length $\xi$ versus normalized eigenvalue for W=0.5, 1, 2, 4 (from top to bottom). Inset: zoom for W=0.5 around the bandwidth center.}
\label{fig_xi}
\end{figure}
\section{Nonlinear Schr\"odinger chain}
We consider the disordered discrete nonlinear Schr\"odinger equation (DNLS) with the Hamiltonian 

\begin{equation}
\mathcal{H}_{D}= \sum_{l} \epsilon_{l} 
|\psi_{l}|^2+\frac{\beta}{2} |\psi_{l}|^{4}
- (\psi_{l+1}\psi_l^*  +\psi_{l+1}^* \psi_l).
\label{RDNLS}
\end{equation}
Here $\psi_{l}$ are complex variables, $l$ are lattice site indices and $\beta \geq 0$ is the nonlinearity strength.  The random on-site energies $\epsilon_{l}$ are chosen uniformly from the interval $\left[-\frac{W}{2},\frac{W}{2}\right]$, with $W$ denoting the disorder strength.  The equations of motion are generated by $\dot{\psi}_{l} = \partial \mathcal{H}_{D}/ \partial (i \psi^{\star}_{l})$:
\begin{equation}
i\dot{\psi_{l}}= \epsilon_{l} \psi_{l}
+\beta |\psi_{l}|^{2}\psi_{l}
-\psi_{l+1} - \psi_{l-1}\;.
\label{RDNLS-EOM}
\end{equation}
Eq.~(\ref{RDNLS-EOM}) conserves the energy (\ref{RDNLS}) and the norm $S= \sum_{l}|\psi_l|^2$.  Varying the norm of an initial wave packet is strictly equivalent to varying $\beta$, therefore we choose $S=1$.  Note that Eq.~(\ref{RDNLS-EOM}) is used to qualitatively describe the evolution of a dilute Bose-Einstein condensate trapped into a deep periodic potential \cite{Billy}, and also the evolution of a light wave in disordered one-dimensional waveguide lattices with cubic Kerr-type nonlinearity (under the paraxial approximation) \cite{Exp2}.

For $\beta=0$, Eq.~(\ref{RDNLS}) with $\psi_{l} = A_{l} \exp(-i\lambda t)$ reduces to the eigenvalue problem 
\begin{equation}
\lambda A_l = \epsilon_l A_l -(A_{l+1} + A_{l-1})\;.
\label{EVequation}
\end{equation} 
The width of the eigenfrequency spectrum $\lambda_{\nu}$ of (\ref{EVequation}) is
$\Delta=W+4$ with $\lambda_{\nu} \in \left[ -2 -\frac{W}{2}, 2 + \frac{W}{2}
\right] $. The normalized eigenvectors $A_{\nu,l}$ ($\sum_l A_{\nu,l}^2=1)$ are the NMs,
and the eigenvalues $\lambda_{\nu}$ are the frequencies of the NMs.  We order the NMs in space by increasing value of the center-of-norm coordinate $X_{\nu}=\sum_l l A_{\nu,l}^2$.

The equations of motion of (\ref{RDNLS}) in normal mode space read
\begin{equation}
i \dot{\phi}_{\nu} = \lambda_{\nu} \phi_{\nu} + \beta \sum_{\nu_1,\nu_2,\nu_3}
I_{\nu,\nu_1,\nu_2,\nu_3} \phi^*_{\nu_1} \phi_{\nu_2} \phi_{\nu_3}\;
\label{NMeq}
\end{equation}
with the overlap integrals
\begin{equation}
I_{\nu,\nu_1,\nu_2,\nu_3} = 
\sum_{l} A_{\nu,l} A_{\nu_1,l} 
A_{\nu_2,l} A_{\nu_3,l}\;.
\label{OVERLAP}
\end{equation}
The variables $\phi_{\nu}$ determine the complex time-dependent amplitudes of the NMs.

\section{Properties of normal modes}
\subsection{Localization length, volume and participation number}
\label{loc_vol}

\begin{figure}
\includegraphics[angle=0,width=0.99\columnwidth]{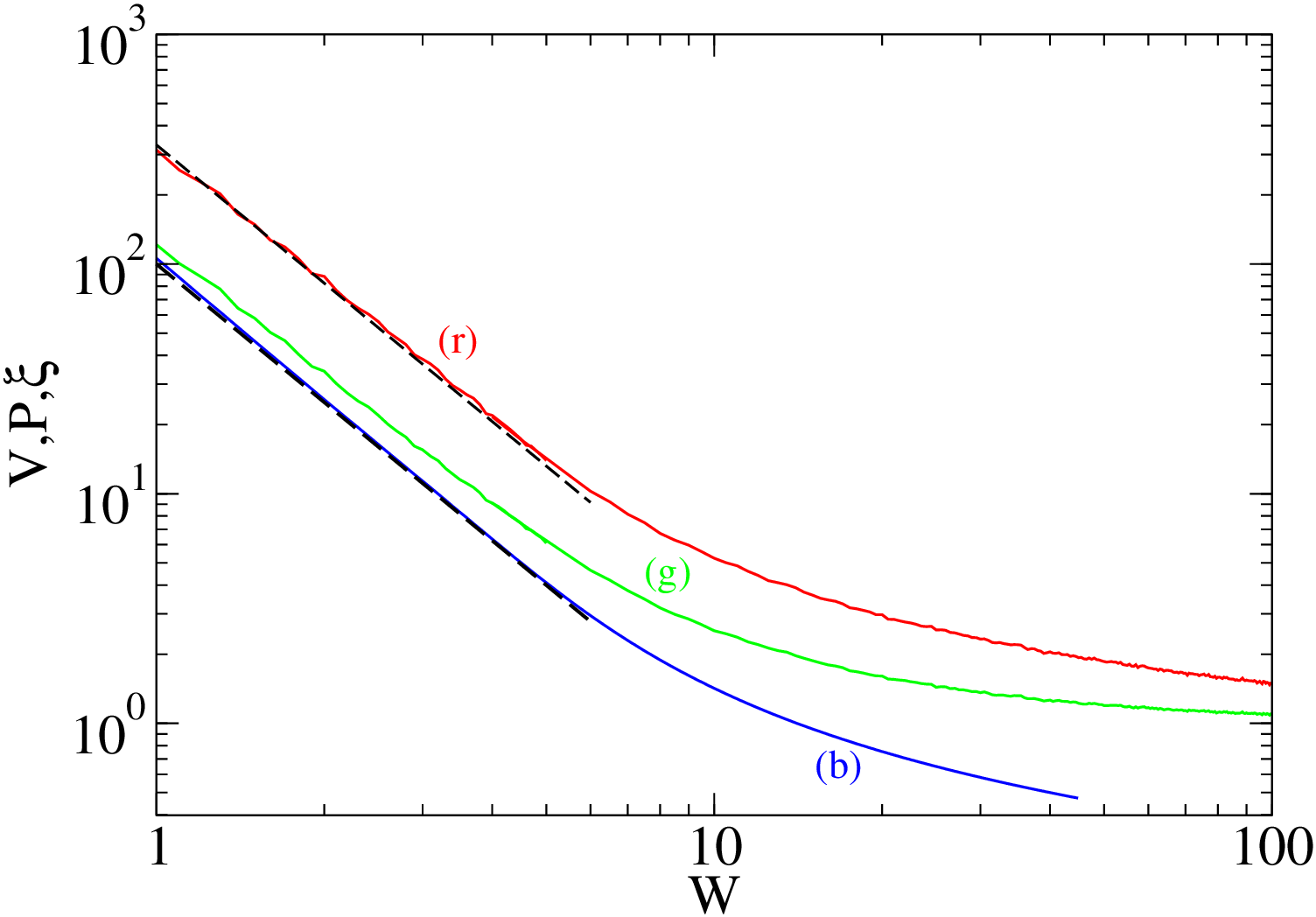}
\caption{(Color online) Average localization volume $V$ [(r)ed], participation number $P$ [(g)reen)] and localization length $\xi$ [(b)lue] of NMs with eigenvalues near the bandwidth center versus strength of disorder $W$. Dashed lines are estimated asymptotics for $V$ and $\xi$ at small disorder strengths, respectively, $330/W^2$ and $100/W^2$. }
\label{fig_L}
\end{figure}

The asymptotic spatial decay of an eigenvector is given by $A_{\nu,l} \sim {\rm e}^{-l/\xi_{\nu}}$, where $\xi_\nu$ is the localization length of a mode $\nu$ with the eigenvalue $\lambda_\nu$. We calculate the average $\xi_\nu$ at a given energy using the standard transfer matrix approach \cite{KRAMER} and show the results in Fig.~\ref{fig_xi}. As expected, the most extended modes correspond to the bandwidth center with $\xi(\lambda=0,W) \approx 100/W^2$ for $W\leq4$ \cite{KRAMER} (see Fig.~\ref{fig_L}). In what follows we refer only to the localization length near the bandwidth center. We also observe 
a small peak at $\lambda =0$ for $W \leq 1$. The smaller the disorder strength, the more pronounced the peak is. For instance for $W=0.5$, the additional peak height
is about $8\%$ of the total value. Its magnitude will not exceed roughly $10\%$ for $W\rightarrow 0$, as first discussed in Refs. \cite{peak}.
The origin of this anomaly is the deviation from single parameter scaling due to the symmetry $A_l(\lambda) = (-1)^l A_l(-\lambda)$ at $W=0$ \cite{sps}. 

Next we estimate the number of NMs which interact with a chosen mode $\nu$. This is assumed to be equivalent to
estimating the number of sites, where the norm density $|A_{\nu,l}|^2$ of the eigenvector is not exponentially small. This number is coined localization volume, $V_\nu$.  $V_\nu$ is related to the localization length $\xi_\nu$,  though quantitatively the two quantities might differ. We consider two ways of estimating this number. A widely used quantity is the participation number $p_{\nu} = 1/\sum_l A_{\nu,l}^4$.
It is a measure of the inhomogeneity of the distribution of eigenvector amplitudes in real space.
Another quantity is the effective distance between the exponential tails of the eigenvector, which is given by
$\sqrt{12 m_2^{(\nu)}}+1$, where $m_2^{(\nu)} = \sum_l (X_{\nu} - l)^2 |A_{\nu,l}|^2$ is the second moment of the norm density distribution.  Both quantities yield the exact width of a flat and compactly distributed norm density distribution. However when fluctuations are included, $p_{\nu}$ will be reduced and underestimates the correct volume, while the effective distance does not. Therefore, we use  $V_\nu=\sqrt{12 m_2^{(\nu)}}+1$ as a measure of the localization volume.

We calculate numerically the average localization volume $V=\overline V_\nu$ and participation number $p=\overline p_\nu$ of NMs as a function of $W$ (see Fig.~\ref{fig_L}). For this purpose we fix the strength of disorder $W$ and take a chain which is much longer than $\xi(W)$. We calculate $V$ and $p$, taking into account only those modes whose eigenvalues are located near the bandwidth center and which are not close to the boundaries of a lattice. Then, we take another realization and repeat the procedure. Finally, we perform the averaging with respect to different disorder realizations. We find that the localization volume $V$ scales on average as $3.3\xi$ for weak disorder (see the dashed lines on Fig.~\ref{fig_L}), and tends to $V=1$ in the limit of strong disorder. We also note that the participation number is almost identical to the localization length for weak disorder, and therefore misses the localization volume
by a factor of three.

Our numerical results indicate that multi-humped NMs are rare and do not affect the statistical results on V. These multi-humped NMs are localized over a number of lattice sites
which can be at far distance from each other in real space (see e.g. \cite{hvyksf10}). If they were statistically relevant, the second moments would be overestimated.
However, we find that in the limit of strong disorder $W \rightarrow \infty$ the localization volume $V \rightarrow 1$, therefore multi-humped states do not significantly contribute
(although we admit that there might be a measurable contribution in this limit of strong disorder). In the limit of weak disorder we observe that $V,P,\xi$ scale in the same
way with the disorder strength, therefore we can exclude any significant statistical relevance of multi-humped NMs in this regime.

\subsection{Overlap integrals}
\label{Overl_int}

We study statistical properties of the {\it absolute values} of the overlap integrals (\ref{OVERLAP}) perturbatively for weak disorder, and numerically
using two different methods. In particular we aim at estimating the average absolute value of these overlap integrals for NMs which are interacting with each
other within the range of one localization volume, in order to exclude statistically irrelevant exponentially weak interactions of distant NMs.
Note, that in following notations, the absolute value is omitted for the sake of simplicity. In order to avoid multiple repetitions in (\ref{OVERLAP}) we use $\nu_3\geq \nu_2\geq \nu_1\geq \nu$. 

{\it Perturbative calculations.}
Let us consider a chain with finite size $N$ and fixed boundary conditions:
\begin{equation}
\lambda A_{l}= W \tilde{\epsilon}_{l} A_{l}
-A_{l+1} - A_{l-1}\;,
\label{oi1}
\end{equation}
$l=1,...,N$, $\psi_0=\psi_{N+1}=0$ and $\tilde{\epsilon}$ are random uncorrelated numbers evenly distributed over the interval $\left[ -1/2,+1/2\right]$.
For $W=0$ the canonical transformation  to standing waves
\begin{equation}
A_l = \sqrt{\frac{2}{N+1}} \sum_{q=1}^N Q_q s_{ql}\;,\; s_{ql} = \sin \left( \frac{\pi q l}{N+1} \right)
\label{oi2}
\end{equation}
yields eigenvectors $A_{q,l} = \sqrt{\frac{2}{N+1}} s_{ql}$.
Eq. (\ref{oi1}) transforms to 
\begin{equation}
\lambda {Q}_q = \lambda_q Q_q + \kappa \sum_{p=1}^N K_{pq} Q_p\;,\;\kappa=\frac{2W}{\sqrt{N+1}}\;,\; 
\label{oi3}
\end{equation}
with the coupling 
\begin{equation}
K_{pq}=\frac{1}{\sqrt{N+1}}\sum_{l=1}^N \tilde{\epsilon}_l s_{ql} s_{pl}
\label{oi4}
\end{equation}
which mixes standing waves with the eigenvalues $\lambda_q=2\cos \left( \pi q/(N+1) \right)$ in the presence of disorder.

The overlap integral
\begin{equation}
I_{q_1,q_2,q_3,q_4} = 
\sum_{l=1}^N A_{q_1,l} A_{q_2,l} 
A_{q_3,l} A_{q_4,l}
\label{oi5}
\end{equation}
at $W=0$ will be zero for all combinations of indices except if a selection rule is satisfied \cite{mishagin08}. It is enough to replace this rule by $\bar{q}_4=\pm q_1 \pm q_2 \pm q_3$. In short we will denote by $\bar{q}_4$ a mode number which satisfies the selection rule for a given triplet of mode numbers $(q_1,q_2,q_3)$. The selection rule applies to $N^3$ overlap integrals $I_0 \sim 1/N$. The other $N^4$ integrals $I_1=0$. Therefore, the average overlap integral becomes $\langle I \rangle(W=0) \sim 1/N^2$.

Let us estimate the corrections to this average when disorder is added. We first consider integrals $I_1$ which were strictly zero at the limit $W=0$.
We perform a perturbation calculation ($W$ small) for a mode $q_4$ such that
$Q_q = Q_q^{(0)} + \kappa Q_q^{(1)} + ...$ with $Q_{q_4}^{(0)}=1$ and $Q_{q\neq q_4}^{(0)}=0$.
Straightforward calculation gives (see also \cite{mvi09})
\begin{equation}
Q_{q \neq q_4}^{(1)} = \frac{K_{q,q_4}}{\lambda_{q_4}-\lambda_q}\;.
\label{oi6}
\end{equation}
Assuming now a triplet of modes $(q_1,q_2,q_3)$ is given, and that $q_4 \neq \bar{q}_4$, the first order nonzero correction to the corresponding overlap integral reads
\begin{equation}
I_{q_1,q_2,q_3,q_4} = \frac{\kappa}{(N+1)^2}\sum_{l=1}^N s_{q_1l}s_{q_2l}s_{q_3l} \sum_{q\neq q_4} 
K_{q,q_4} \frac{s_{ql}}{\lambda_{q_4}- \lambda_q}\;.
\label{oi7}
\end{equation}
We started with $q_4 \neq \bar{q}_4$, but in the presence of disorder the mode with number $q=\bar{q}_4$ will become excited. Therefore, after summation over $l$ in (\ref{oi7}) we find
\begin{equation}
I_{q_1,q_2,q_3,q_4} = \frac{\kappa}{(N+1)}  \frac{K_{\bar{q}_4,q_4}}{\lambda_{q_4}- \lambda_{\bar{q}_4}}\;.
\label{oi8}
\end{equation}
Note that the indices $(q_1,q_2,q_3)$ are implictely hidden in the quantity $\bar{q}_4$.
In order to estimate the average, we have to take the {\it absolute value} of (\ref{oi8}), to sum over each index $q_i$, $i=1,2,3,4$ and each time to divide by $N$.
Let us perform the averaging over $q_4$. The denominator $\lambda_{q_4}- \lambda_{\bar{q}_4}$ will become of the order of $1/N$ when $q_4$ is close
to $\bar{q}_4$. Replacing the sum by an integral, we estimate
\begin{equation}
\frac{1}{N}\sum_{q_4 \neq \bar{q}_4} |I_{q_1,q_2,q_3,q_4}| \sim \frac{\kappa}{N+1} \ln (N) \frac{|K_{\bar{q}_4,\bar{q}_4}|}{|\sin (\pi \bar{q}_4/N)|}\;.
\label{oi9}
\end{equation}
Since the disorder average $\langle K_{p,q} \rangle = 0$ and its variance is finite (i.e. not depending on $N$) the final averaging over $q_1,q_2,q_3$ yields
\begin{equation}
\langle I_1 \rangle \sim \kappa \ln (N) / N.
\label{oi9-2}
\end{equation} 
The overlap integrals $I_0$ for $q_4=\bar{q}_4$ were of the order of $1/N$ for $W=0$. It is straightforward to obtain that the disorder induced correction will be of the same order as (\ref{oi9-2}), which is still smaller than the unperturbed value.

Thus the average value of $\langle I \rangle$ up to the first order of perturbation in $W$ is given by (remember that 
$\kappa=\frac{2W}{\sqrt{N+1}}$)
\begin{equation}
\langle I \rangle\sim1/N^2+aW\ln (N) / N^{3/2},
\label{eq_I_full}
\end{equation}
where $a$ is some constant independent on the system's parameters. One can conclude from Eq.~(\ref{eq_I_full}), that for small enough $W$ such that $W < (\sqrt{N} \ln (N))^{-1}$, the first term prevails and the total average integral is $\langle I \rangle \sim 1/N^2$. In the opposite case, when $W > (\sqrt{N} \ln (N))^{-1}$, the second term in Eq.~(\ref{eq_I_full}) dominates and, as a result, we get $\langle I \rangle \sim W N^{-3/2} \ln (N)$.

Note that the perturbed eigenvectors $A_{ql} = A^{(0)}_{ql} + \kappa A^{(1)}_{ql} + ...$, given by $A^{(1)}_{ql} = \sum_{p\neq q}  K_{q,p} \frac{s_{ql}}{\lambda_{q}- \lambda_p}$ do not yield logarithmic divergence, since - at variance to the overlap integrals - no absolute values are taken, and the two logarithms obtained from integrating to the left and right of $q$ are cancelling each other due to opposite signs.

For $W \rightarrow \infty$ the NM eigenvectors become single site profile, and the overlap integrals tend to zero. Therefore, for a given size $N$,
the average overlap integral will start to increase with $W$ for $W > (\sqrt{N} \ln (N))^{-1}$, reach a maximum at $W_{max}$, and decay down to zero for infinitely strong disorder.
It is reasonable to assume that the localization volume $\xi(W_{max}) \sim N$. In that case for small values of $W$ we obtain 
\begin{equation}
\langle I \rangle \sim \frac{\ln (V)}{V^2} \sim - W^4 \ln (W) \;,
\label{oi10}
\end{equation}
which is an estimate of the interaction strength of NMs within the spatial range of one localization volume.

\begin{figure}
\vspace*{0.5cm}
\includegraphics[angle=0,width=0.99\columnwidth]{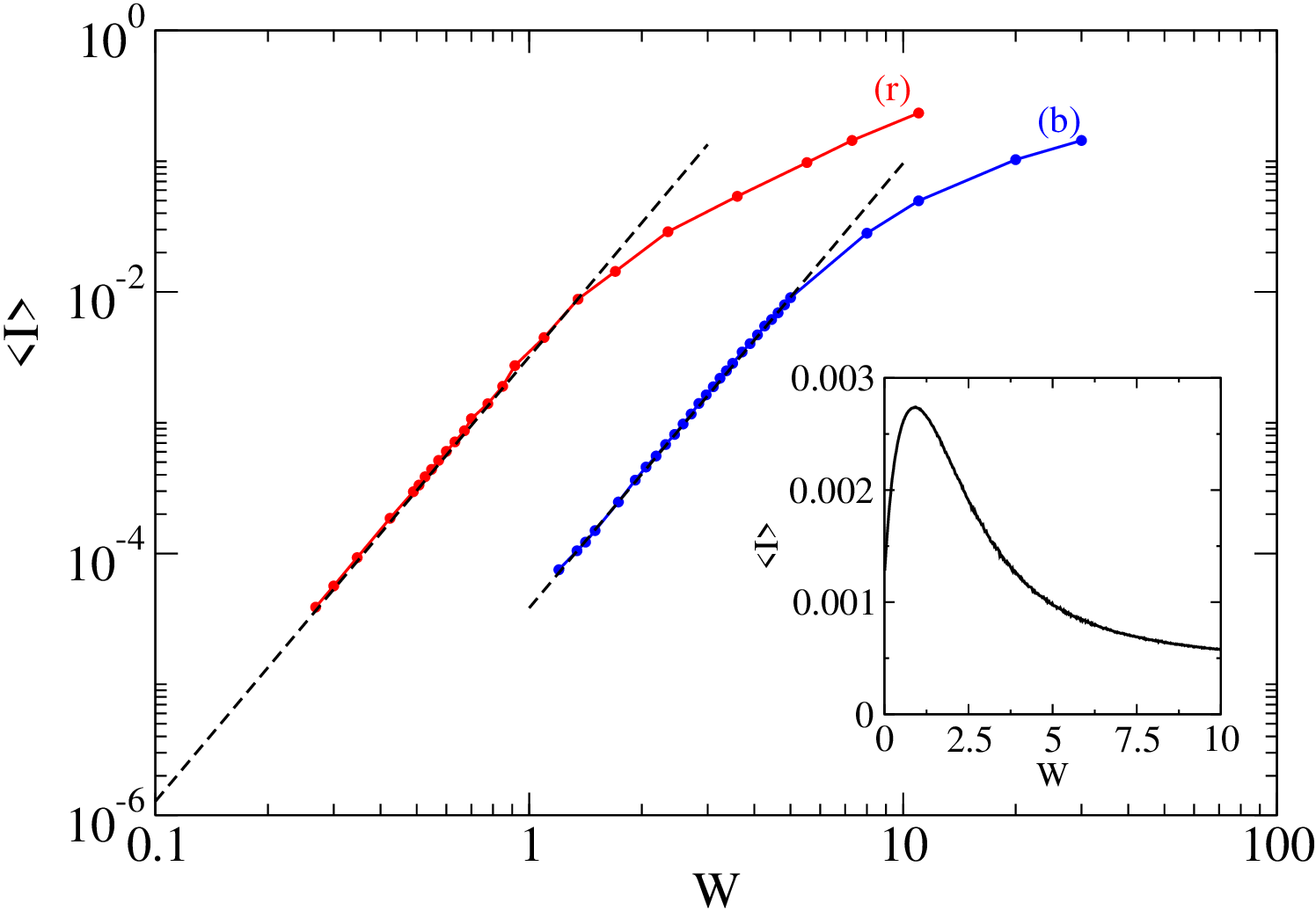}
\caption{(Color online) Average integrals $\langle I \rangle$ versus strength of disorder $W$ using method I
[(r)ed] and method II [(b)lue]. Dashed fitting lines are: $3.4\cdot 10^{-3} W^{3.4}$ (upper line) and $3.84\cdot 10^{-5} W^{3.4}$ (lower line).  Inset:  $\langle I \rangle$ versus $W$ for $N=40$. For the averaging, 400 disorder realizations were used.}
\label{fig_I_aver_vs_W}
\end{figure}

{\it Numerical calculations. Method I.}
\label{Overl_int_mI}
We fix a chain size $N$ and calculate the average value of the overlap integrals taking into account {\it all} integrals (\ref{OVERLAP}). Then, we take another realization and repeat the procedure. Finally, we perform the averaging with respect to different disorder realizations. Each averaged integral $\langle I \rangle$ is a function of $W$. As derived in the above perturbation approach, it has a maximum value $\langle I \rangle(W_{max})$ at a certain $W_{max}$, as is shown in the inset of Fig.~\ref{fig_I_aver_vs_W}
for $N=40$. 
Now we vary the chain size $N$, and repeat the procedure.
In Fig.~\ref{fig_I_aver_vs_W} we plot the maximum values of  $\langle I \rangle(W_{max})$ as a function of $W_{max}$ (red curve). We find that $\xi(W_{max}) / N \approx 8/3$, as expected in the above perturbation approach. We also find that for large $N$ the data can be fitted with the power law  
$\langle I \rangle \approx 0.0034 W^{\alpha}$, with $\alpha=3.40\pm 0.02$ (see Fig.~\ref{fig_I_aver_vs_W}). The fit was done using different number of numerical points (from 3 to 14) starting from the smallest W. In all cases the RMS relative error was better than $10^{-3}$. We expect that this method will overestimate the corresponding  prefactors. This is due to the fact that $N \approx 0.38 \xi(W_{max}) $ and therefore states overlap more strongly than in an extended system, as seen in the next method. 

{\it Method II.}
We fix the strength of disorder $W$ and choose a chain size $N  \gg \xi(W)$. We  select a middle part of a smaller size (core) of the width $L$, and do not consider the edges in order
to avoid boundary effects. We use only modes within the core ($\nu=1..L$). For each mode $\nu$ we calculate its localization volume $V_\nu$.  Now we consider only NMs which happen to reside in a corresponding neighborhood, i.e.  we select $V_\nu/2$ modes from the right and left (in case $V_\nu$ is odd, one mode is randomly taken from left or right in addition). Therefore, we have defined a subset of NMs which interact with the $\nu$th NM. We calculate all overlap integrals for this subset. Then, we move on to the next reference mode from the core. This procedure is performed for all NMs from the core, for many realizations. For small $W$ the data can be fitted with the power law $ \langle I \rangle \approx 3.84\cdot 10^{-5} W^{3.4}$ (see  Fig.~\ref{fig_I_aver_vs_W}) . Note that both methods yield the same exponents. Note also that  we lack more data to distinguish between the numerically found law $W^{3.4}$ and the perturbation result $- W^4 \ln (W)$.
%
\section{Frequency scales}
\label{Sec_d_sp}

There are two frequency scales set by the linear equations (\ref{EVequation}): the average spacing $d$ of NMs within the range of a localization volume and the width of the spectrum $\Delta$ \cite{fks09,skfk09,F10}. The two scales $ d \leq \Delta$ determine the packet evolution details in the presence of nonlinearity. 
In order to calculate the average spacing and its distributions numerically,
%
%
%
we fix the strength of disorder $W$ and take a chain which is much longer than $V(W)$. We select a middle part of a smaller size (core), and do not consider the edges. For each mode $\nu$ within a core we form its subspace which consists of those modes which live in its localization volume $V_\nu$ [see Sec.~\ref{loc_vol},\ref{Overl_int} for details]. We take the eigenvalues of these modes (including the eigenvalue of $\nu$-th mode), sort them and compute absolute values of spacings between them. Then, we proceed to the next reference mode from the core. This procedure is performed for all NMs from the core and for many realizations, such that we end up with a large number of spacings (usually of the order of $10^6$)
\cite{methodII}.
%
%
%
%
\begin{figure}
\includegraphics[angle=0,width=0.9\columnwidth]{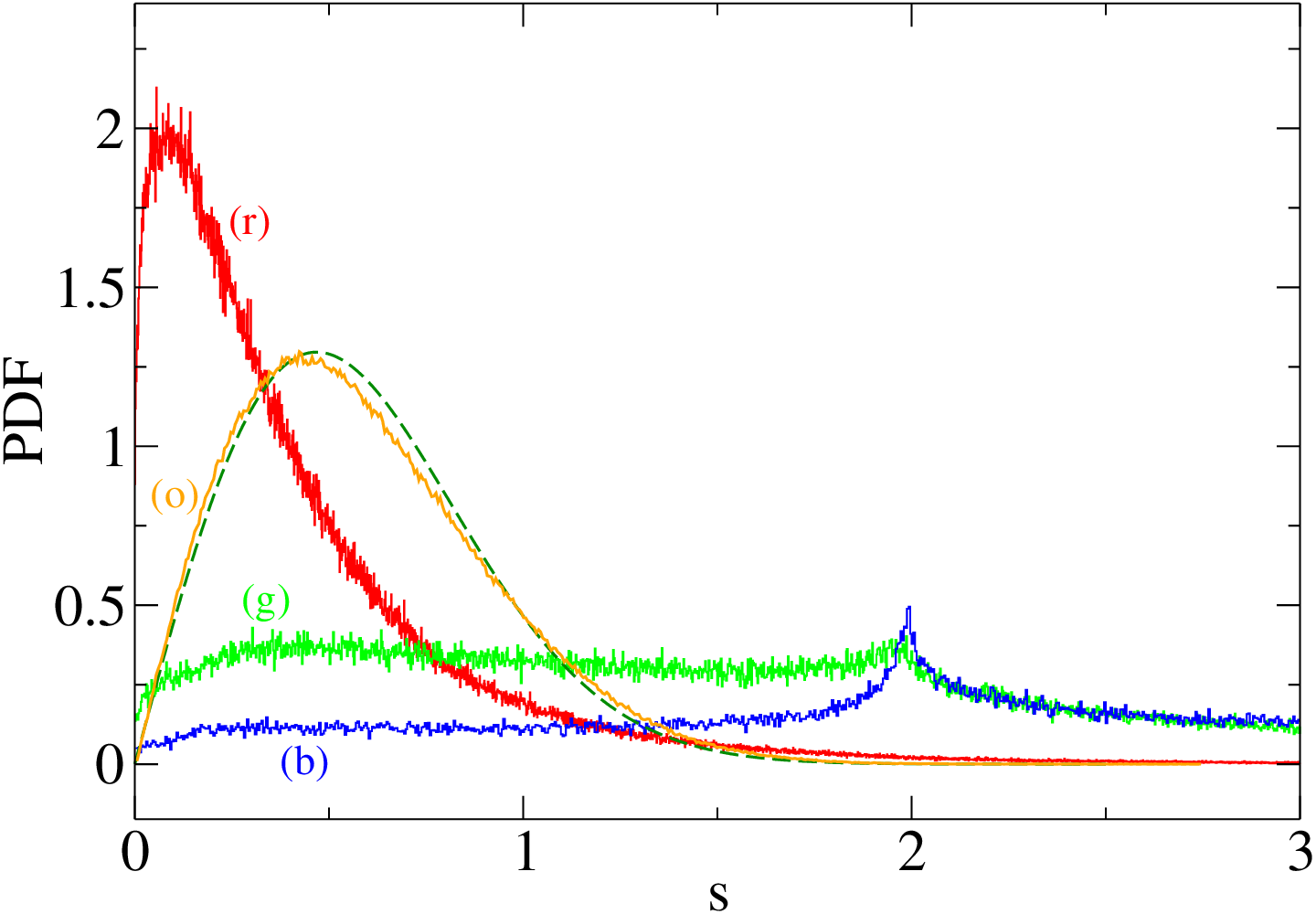}
\caption{(Color online) PDFs of eigenvalue spacings $s$ for W=4,10,20 [(r)ed,(g)reen,(b)lue]. (O)range curve: PDF for W=4 and a short chain with N=10. Dashed curve: Wigner-Dyson distribution with the average spacing $d\approx 0.59$ [see Eq.~(\ref{WgD})].}
\vspace*{0.8cm}
\label{fig_PDF_dL_dist}
\end{figure}
\begin{figure}[!]
\includegraphics[angle=0,width=0.9\columnwidth]{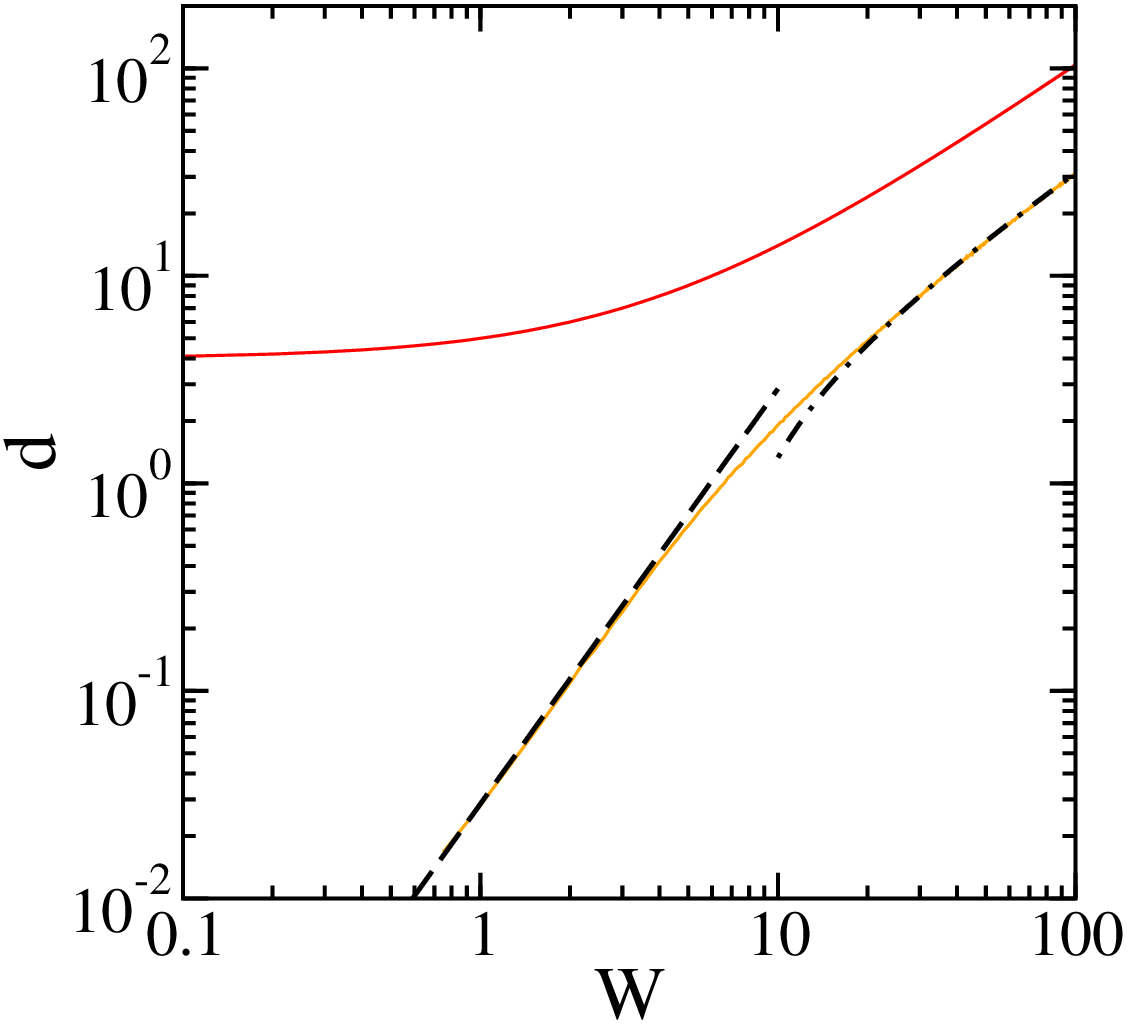}
\caption{(Color online) Orange (light gray) curve: numerically calculated average spacing $d$ versus strength of disorder $W$ (see Sec.~\ref{Sec_d_sp}). Red (dark gray) curve: width of the spectrum $\Delta$. Dashed curve: the fit of $d$ in the limit of $W \rightarrow 0$ by $d=W^2/37$. Dashed-doted curve: the fit of $d$ in the limit of $W \rightarrow \infty$ by $d=W/3-2$.}
\label{fig_diagram_log}
\end{figure}

Typical probability density functions (PDF) of the spacings $s$ are shown in Fig.~\ref{fig_PDF_dL_dist}. 
For strong disorder $W\gg1$ the relative contribution of small spacings to the PDF becomes smaller. The reason is that the localization volume tends to one and for each reference mode $\nu$ we take into account only a single neighboring mode. As a result, spacings between eigenvalues increase. 

We also note that the computed PDFs are far from following a Wigner-Dyson distribution
\begin{equation}
P(s)=\dfrac{\pi s}{2 d^2}\cdot e^{-\dfrac{\pi s^2}{4d^2}}\;.
\label{WgD}
\end{equation}
Especially for small $W$, such a distribution could be expected due to large localization lengths. However we find systematic deviations towards a Poisson distribution with an enhancement of the probability density at small spacings. This is due to the fact that NMs overlap in general only partially in real space. The Wigner-Dyson distribution is recovered only for very short chains, when $N<\xi(W)$ [see Fig.~\ref{fig_PDF_dL_dist}]. In this case, all eigenmodes occupy the same volume, and level repulsion is recovered as expected. Interestingly, similar level repulsion occurs in short resonators in one-dimensional random lasers. In that case the PDF of spacings between frequencies of the neighboring lasing modes tends also to the Wigner-Dyson distribution \cite{zds09}. 

In Fig.~\ref{fig_diagram_log} we plot the result for the average spacing $d$. In the limit of small $W$ we estimate the average spacing as $d=\Delta/V\propto W^2$. For large disorder strength $W \gg 1$ the localization volume tends to one. Thus, only two modes form a subspace of each reference mode $\nu$ (one of which is the mode $\nu$ itself). Therefore, the spacing can be calculated by considering two numbers (emulating two corresponding eigenvalues) which are randomly distributed within the width of the spectrum $\Delta$. The average distance between these numbers $x$ and $y$, assuming that $x\ge y$ is 
\begin{equation}
\label{asym_W_large}
d=\dfrac{1}{N}\int_0^\Delta dx \int_0^x (x-y) dy,\,\,\,\,N=\int_0^\Delta dx \int_0^x dy.
\end{equation}
It follows $d = \Delta/3$. In Fig.~\ref{fig_diagram_log} the two theoretical estimates are shown to be close to the numerical data.

\begin{figure}
\includegraphics[angle=0,width=0.9\columnwidth]{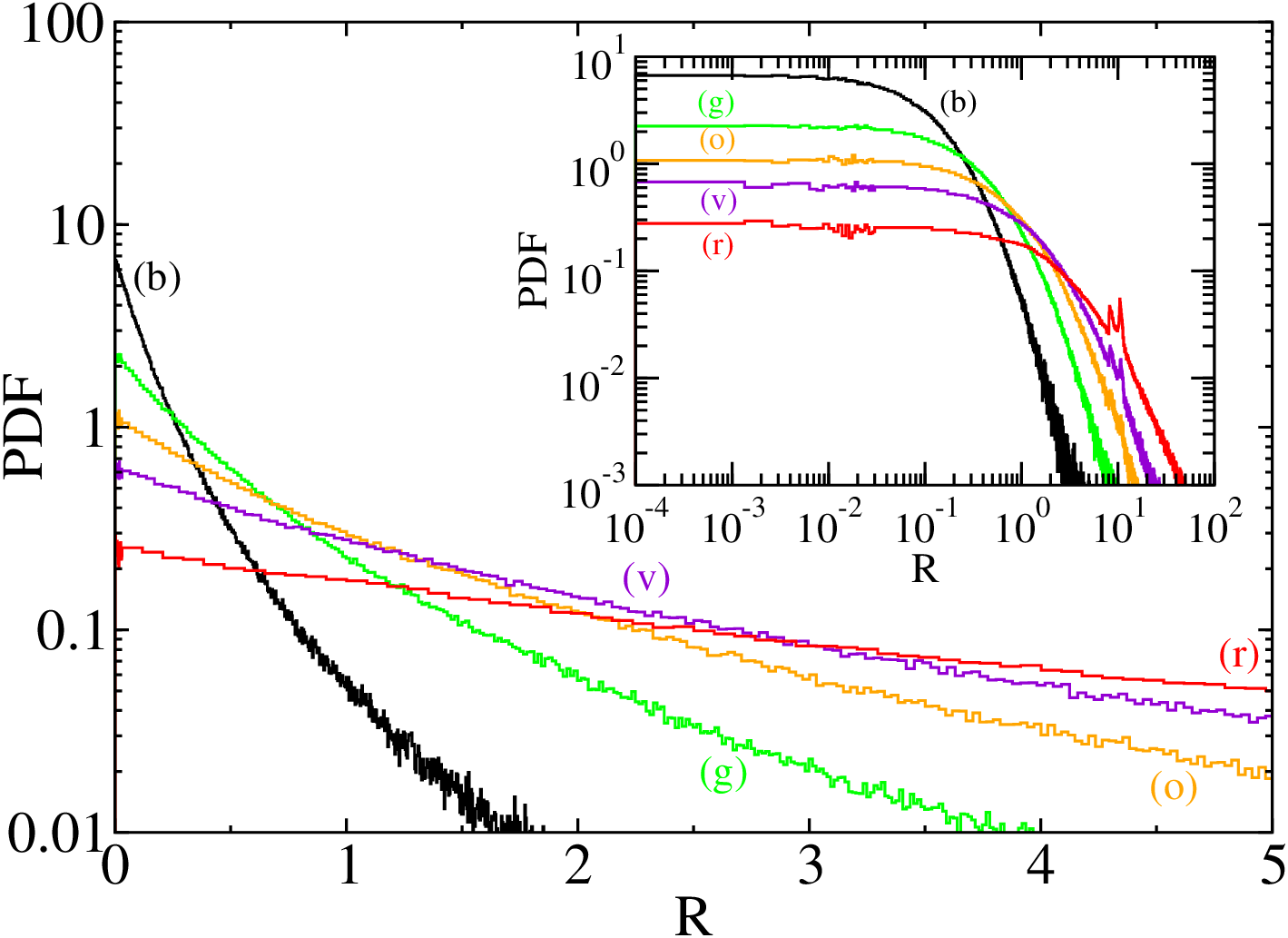}
\caption{(Color online) Probability densities $\mathcal{W}(R_{\nu,\vec{\mu}_0})$ of resonant NMs in linear-log scale (main figure) and log-log scale (inset). Disorder strength $W=4,6,8,10,15$ [(b)lack, (g)reen, (o)range, (v)iolet, (r)ed].  }
\label{fig_PDF_R_full_st}
\vspace*{0.5cm}
\end{figure}
\begin{figure}
\includegraphics[angle=0,width=0.9\columnwidth]{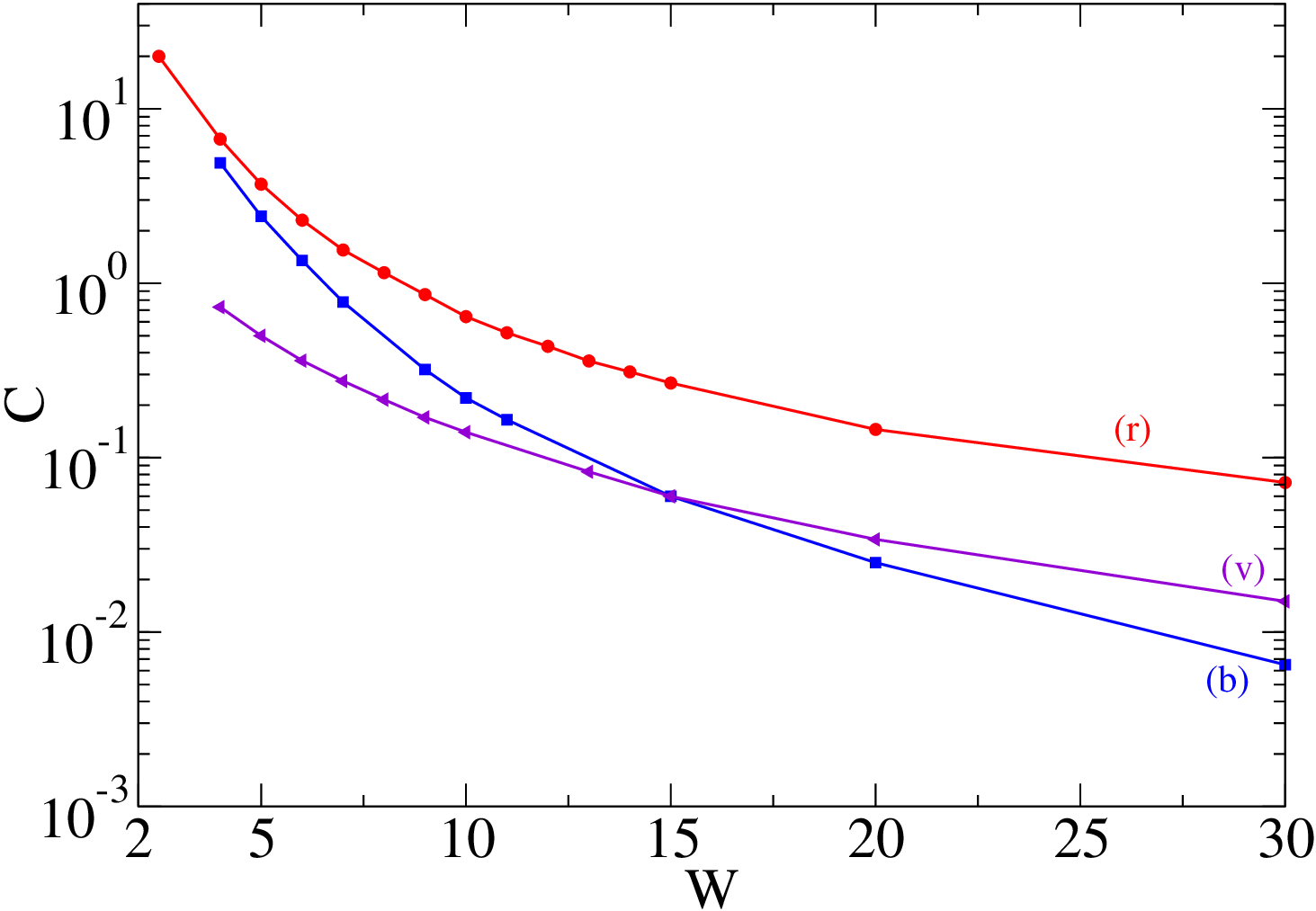}
\caption{(Color online) The constant $C$ as a function of $W$ in linear-log scale when taking into account contributions from i) all combinations [(r)ed]; ii) quadruplets  [(b)lue], and iii) triplets [(v)iolet].}
\label{fig_R0_vs_W}
\end{figure}
\section{Resonances}
\label{res_int}
When a wave packet spreads, its size grows, and the norm density inside the packet drops. Previous studies indicate that this process of spreading is due to resonances in mode-mode interactions. These resonances lead to chaotic dynamics inside the packet, and to a subsequent incoherent spreading. Let us estimate the number of resonant modes in the packet.  Excluding secular interactions, the amplitude of a NM with
$|\phi_{\nu}|^2 = n_{\nu}$ is modified by a set of three other modes  $\vec{\mu}\equiv (\mu_1,\mu_2,\mu_3)$ in first order in $\beta$ as (\ref{NMeq}) (see  \cite{skfk09})
\begin{eqnarray}
\label{PERT1}
|\phi_{\nu}^{(1)}| &=& \beta \sqrt{n_{\mu_1}n_{\mu_2}n_{\mu_3}}
R_{\nu,\vec{\mu}}^{-1}\;,\;
\\ 
R_{\nu,\vec{\mu}} &\sim&
\left|\frac{\lambda_{\nu}+\lambda_{\mu_1}-\lambda_{\mu_2}-\lambda_{\mu_3}}{I_{\nu,\mu_1,\mu_2,\mu_3}}\right| \;.
\label{Rmu_nu}
\end{eqnarray}
The perturbation approach breaks down and resonances set in when $\sqrt{n_{\nu}}< |\phi_{\nu}^{(1)}|$.  Since all considered NMs belong to the packet, we assume their norms to be equal to $n$.  

We perform a statistical numerical analysis by computing the PDF of $R_{\nu,\vec{\mu}}$. For a given NM $\nu$ we obtain $ R_{\nu,\vec{\mu}_0} = \min_{\vec{\mu} } R_{\nu,\vec{\mu}}$. Collecting $R_{\nu,\vec{\mu}_0}$ for many $\nu$ and many disorder realizations, we find the probability density distribution
$\mathcal{W}(R_{\nu,\vec{\mu}_0})$. We also analyze separate contributions from three different types of interactions, namely from quadruplets (all four modes are different), triplets (only three of four modes are different) and pairs (only two different modes participate in the interaction). For quadruplets all indices in (\ref{Rmu_nu}) should be different, i.e.
$\mu_1 \neq \nu,\,\mu_2 \neq \nu,\,\mu_3 \neq \nu,\,\mu_1 \neq \mu_2,\,\mu_1 \neq \nu,\,\mu_1 \neq \mu_3,
\,\mu_2 \neq \mu_3$. For triplets either $\mu_1=\nu$ such that 
\begin{eqnarray}
R_{\nu,\vec{\mu}} &\sim&
\left|\frac{\lambda_{\mu_2}-2\lambda_{\nu}+\lambda_{\mu_3}}{I_{\nu,\nu,\mu_2,\mu_3}}\right| \;,
\label{Rmu_nu_triplet1}
\end{eqnarray}
or $\mu_2=\mu_3$ with
\begin{eqnarray}
R_{\nu,\vec{\mu}} &\sim&
\left|\frac{\lambda_{\nu}-2\lambda_{\mu_2}+\lambda_{\mu_1}}{I_{\nu,\mu_1,\mu_2,\mu_2}}\right| \;.
\label{Rmu_nu_triplet2}
\end{eqnarray}
The remaining cases form the subset of pairs. 

The probability densities $\mathcal{W}(R_{\nu,\vec{\mu}_0})$ of NMs being resonant when taking into account all contributions are shown in Fig.~\ref{fig_PDF_R_full_st}. The main result is that $\mathcal{W}(R_{\nu,\vec{\mu}_0} \rightarrow 0) \rightarrow C(W) \neq 0$. The constant $C$ drops with increasing disorder strength $W$ (see Fig.~\ref{fig_R0_vs_W}). We also calculate $C$ by taking into account only quadruplets and triplets (see Fig.~\ref{fig_R0_vs_W}). We find that for weak disorder the quadruplet contributions are the dominant ones, while for strong disorder their contribution diminishes as compared to the triplet contribution.

For small $R$ the probability densities $\mathcal{W}(R)$ can be approximated as
\begin{eqnarray}
W(R)\approx C(W)e^{-C(W)R}.
\label{WR_appxmt}
\end{eqnarray}
The probability $\mathcal{P}$ for a mode, which is excited to a norm $n$, to be resonant at a given value of the interaction parameter $\beta$ is given by
\begin{equation}
\mathcal{P} = \int_0^{\beta n} \mathcal{W}(R) {\rm d}R \approx 1-e^{-C \beta n} \;.
\label{resprob}
\end{equation} 
\section{Discussion}

We have studied statistical properties of eigenvalues and eigenvectors of waves in disordered one-dimensional systems as a function of the disorder strength. We estimated the localization volume of a mode which defines the number of interacting partner modes.  We obtained the dependence on the disorder strength of the overlap integrals which determine the interaction strength. We analyzed the statistics of level spacings of normal modes within one localization volume. Finally, we obtained distribution functions for resonance probabilities of normal modes interacting in the presence of nonlinearity. Let us discuss some of the consequences of our findings.

\subsection{Overlap integrals}

In order to estimate the {\it absolute value} of the overlap integral (\ref{OVERLAP}) for modes within one localization volume for weak disorder, 
Shepelyansky \cite{Shep94} and Imry \cite{Imry95} assumed that the sum extends roughly over the localization volume $V$, with each term in the sum $A_{\nu,l} A_{\nu_1,l} A_{\nu_2,l} A_{\nu_3,l}$ having a random sign. The absolute value of the eigenvector is of the order of $1/V^{1/2}$ due to normalization. Then (\ref{OVERLAP}) can be evaluated using
the central limit theorem, for which the average absolute total value $\langle I \rangle_{rp} \sim V^{-3/2}\sim W^{3}$. Our numerical finding $\langle I \rangle\sim W^{3.4}$ 
clearly rules out the random sign resut $W^3$. 
As shown in the perturbation calculation in section III, the reason for the random sign failure is that NMs are similar to plane waves with definite phases on each lattice site
(inside the localization volume). These phases enforce selection rules, which become strict in the very limit $W=0$.

While we can now exclude the random sign result $W^3$, we can not tell whether the numerical estimate $\langle I \rangle \sim W^{3.4}$ is correct, or the
perturbation result $\langle I \rangle \sim - W^4 \ln (W)$ will set in for small enough $W$. 
Ponomarev and Silvestrov \cite{ponomarev97} have also stressed the importance of phase correlations in Eq.~(\ref{OVERLAP}). 
A numerical calculation of the average of the squared overlap integral was performed by Frahm et al \cite{frahm95} for $1.4 < W < 4$ yielding $\langle I \rangle\sim W^{3.3}$, in a good agreement with our numerical data.

The random sign estimate $W^3$ was taken to predict a strong increase of the  localization length of two  interacting particles in a one-dimensional random quantum chain \cite{Shep94,Imry95}. The two particle localization volume $V_2$, within a renormalization group approach, is given by $V_2/V \sim  \langle I \rangle^2 V^4$ where $V$ is the single particle localization volume. For the random phase result, this yields $V_2 \sim V^2$ \cite{Shep94,Imry95,oppen96}. We can clearly rule out such an outcome. Instead, we expect either $V_2 \sim V^{1.6}$ (numerical data) or $V_2 \sim V \ln^2 V$ (perturbation approach), which give a much weaker effect. These controversies call for more detailed investigations.

\subsection{Asymptotic spreading of wave packets in nonlinear chains}

According to a recent analysis of the spreading scenaria of wave packets \cite{F10}, the only scale which separates different dynamical spreading regimes  
is the average spacing $d$. Therefore,  the constant $C$ from the previous section is inversely
proportional to the mean level spacing:
\begin{equation}
C \sim \frac{1}{d}\;.\label{disc1}
\end{equation}

Following the theory developed in \cite{fks09,skfk09,F10} for the asymptotic spereading, an exterior mode ${\phi}_{\mu}$ which is heated up by the packet obey the following evolution equation in accordance with (\ref{NMeq}) 
\begin{equation}
i \dot{\phi}_{\mu} \approx \lambda_{\mu} \phi_{\mu} + \beta \langle I \rangle V^3 \mathcal{P}(\beta n) n^{3/2} f(t),
\end{equation}
where $\langle f(t) f(t') \rangle = \delta(t-t')$ ensures that $f(t)$ has a continuous frequency spectrum. Note, that here we also introduce the contribution of the overlap integrals estimated as $ \langle I \rangle V^3$. Repeating the previous derivations  \cite{fks09,skfk09,F10}, we finally get the following expression for the asymptotic growth of the second moment of spreading wave packets in nolinear chains
\begin{equation}
m_2 \sim \beta^{4/3} V^{8/3} \langle I \rangle^{2/3} t^{1/3}\;.
\label{spreading1}
\end{equation}
From our numerical data for weak disorder it follows  $m_2 \sim W^{-3.07} \beta^{4/3} t^{1/3}$, while the perturbation approach yields $m_2 \sim W^{-8/3} (-\ln W)^{2/3} \beta^{4/3} t^{1/3}$. The prefactor dependence of $W$ is another intriguing test which awaits numerical verification.

\section{Conclusion}

In conclusion, we performed a statistical analysis and calculated the average localization volume occupied by an eigenmode as a function of disorder strength which determines the average number of a nonexponentially interacting eigenmodes. Then, we calculated the frequency spacings of the normal modes which happen to interact in a nonexponentially weak way and their distributions and the average numerically. This result  is very important for the classification of different regimes of wave packet spreading in the presence of nonlinearity.  We also studied statistical properties of the overlap integrals which determine the coupling strength between the interacting modes and, thus, influence properties of spreading. Finally, we estimated the number of resonant modes in the packet and proved that the most significant contribution to the spreading comes from the quadruplet and triplet resonances for small to moderate values of disorder strengths, and from triplets for the case of large disorder.

\section{Acknowledgements} The authors thank I. Aleiner, B. Altshuler, J. Bodyfelt, R. Khomeriki, T. Lapteva, N. Li and Ch. Skokos for useful discussions.

{}

\begin{thebibliography}{}
\bibitem{methodII}
We tested another method by
computing all overlap integrals $I_{\nu,\mu} = \sum_{l} A_{\nu,l}^2 A_{\mu,l}^2$ within a core with the reference mode $\nu$. Then, we choose $V_\nu$ modes  (together with the $\nu$-th mode) which give the largest overlap integrals and compute spacings between their eigenvalues as describe above. The final results are almost identical to the ones discussed in the main text.
\end{thebibliography}
\end{document}